\documentstyle[11pt,newpasp,twoside,epsf]{article}

\newcommand{\msun}{$M_\odot$}
\newcommand{\etal}{{et~al.}}
\newcommand{\mhi}{$M_{HI}$}
\newcommand{\HI}{\protect\ion{H}{I}}
\newcommand{\kms}{km~s$^{-1}$}

\begin{document}
\title{An \HI\ Census of Loose Groups of Galaxies}
\author{D.J. Pisano}
\affil{NSF Distinguished International Postdoctoral Research Fellow, ATNF, 
P.O. Box 76, Epping NSW 1710, Australia}
\author{David G. Barnes}
\affil{School of Physics, Univ. of Melbourne, Victoria 3010, Australia}
\author{Brad K. Gibson}
\affil{Swinburne University of Technology, Centre for Astrophysics and 
Supercomputing, Mail \# 31, , P.O. Box 218, Hawthorn, Victoria 3122, Australia}
\author{Lister Staveley-Smith}
\affil{ATNF, P.O. Box 76, Epping NSW 1710, Australia}
\author{Ken C. Freeman}
\affil{RSAA, Mount Stromlo Observatory, Cotter Road, Weston, ACT 2611, 
Australia}

\begin{abstract}
We present results from our Parkes Multibeam \HI\ survey of 3 loose
groups of galaxies that are analogous to the Local Group.  This is a 
survey of groups containing only spiral galaxies with mean
separations of a few hundred kpc, and total areas of approximately 1 
Mpc$^2$; groups similar to our own Local Group.  We present a
census of the \HI-rich objects in these groups down to a 1$\sigma$ \mhi\
sensitivity $\sim$7$\times$10$^5$\msun, as well as the detailed properties of 
these detections from follow-up 
Compact Array observations.  We found 7 new \HI-rich members in the 3
groups, all of which have stellar counterparts and are, therefore, typical
dwarf galaxies.  The ratio of low-mass to high-mass gas-rich galaxies in these 
groups is less than in the Local Group meaning that the ``missing satellite'' 
problem is not unique.  No high-velocity cloud analogs were found in any of 
the groups.  If HVCs in these groups are the same as in the Local Group, this 
implies that HVCs must be located within $\sim$300-400 kpc of the Milky Way.  
\end{abstract}

\section{Introduction}

Loose groups of galaxies are collections of a few (2-3) large galaxies and
tens of smaller galaxies.  They are the most diffuse components of structure
in the universe, yet they are relatively understudied despite their importance.
About 60\% of galaxies reside in groups (Tully 1987) including the Milky Way
which is part of the Local Group of galaxies.  Loose groups are possibly still
forming (Zabludoff \& Mulchaey 1998), and may even be the site of ongoing 
galaxy formation as traced by the high-velocity clouds (HVCs; e.g. Blitz \etal\
1999).  Measuring the distribution of galaxy masses in groups also provides a 
useful constraint on models of galaxy formation.  To learn more about groups
of galaxies similar to our own, to constrain the nature of HVCs, and to 
learn more about galaxy and structure formation, we have commenced a survey 
of loose groups of galaxies.

\section{Observations}

Our survey examined analogs to the Local Group:  spiral-rich loose
groups without any large elliptical or lenticular galaxies.  We selected 
five such groups from the LGG catalog of Garcia (1993).  A sixth group was 
selected from the HIPASS group catalog (Stevens, 2003, private communication)
and was not previously identified optically.  Assuming H$_0$ = 65 km/s/Mpc, 
the groups lie between 10.6 - 13.4 Mpc.  We observed a projected area of 
1 - 1.7 Mpc$^2$ centered on each group using the Parkes Multibeam receiver.  
The observations involved scanning the instrument in a ``basket-weave'' pattern
in right ascension and declination multiple times until an RMS sensitivity of
$\sim$7$\times$10$^5$\msun\ per 3.3 km/s was reached.  The total velocity
coverage of the observations was either 1700 or 3400 km/s (depending on the
group) with a velocity resolution of 1.65 km/s or 3.3 km/s.  The linear
resolution of the Parkes observations was $\sim$50 kpc. 

The data were reduced and gridded into cubes using the ATNF {\bf livedata} and 
{\bf gridzilla} packages in {\bf aips++}.  The final cubes were searched by
eye by three people for detections.  Fake sources were inserted into these 
cubes to assess the reliability and completeness of the search.  If a source
was found by at least 2 people, it was considered to be a detection.  Based
on this analysis, our detection algorithm was essentially 100\% complete down 
to the 10$\sigma$ level.
 
All our detections, not just the new ones, were re-observed with the ATCA.  
The ATCA observations not only served to 
confirm the detections, but the higher spatial resolution, $\sim$4 kpc, 
allowed us to uniquely identify \HI\ detections with optical counterparts.  
Furthermore, as the ATCA observations had about the same sensitivity as the
Parkes cubes, we were able to search for \HI\ clouds which were previously
unresolved in the Parkes data.  

The combination of the large area observed, high velocity resolution, and 
extremely sensitive observations of multiple loose groups makes this survey 
uniquely tuned to get a census of the \HI\ content of loose groups of galaxies 
and to strongly constrain the origin of HVCs.  We will address these topics
using the first 3 groups studied in our survey:  LGG 93, LGG 180, and LGG 
478.  

\section{The \HI\ Content of Loose Groups of Galaxies}

Cold dark matter simulations of the formation of the Local Group of 
galaxies (e.g. Klypin \etal\ 1999, Moore \etal\ 1999) reveal an extremely 
large number of small dark matter halos ($\sim$300) compared to the number of 
known luminous satellite galaxies ($\sim$20).  This is the so-called 
``missing satellite'' problem.  While invoking a different type of dark 
matter, such as warm dark matter, or invoking feedback processes to suppress
the collapse of baryons in dark halos can solve this problem, it is worth
asking if this problem exists at the same level in groups similar to the 
Local Group.  

Our survey detected all the optically identified galaxies in the 3 groups.  In
addition, we detected 4 new group members in LGG 93, 2 new group members in 
LGG 180, and 1 new group member in LGG 478 (along with 7 background galaxies). 
Only 4 of these new detections were not evident in the HIPASS data
(Barnes \etal\ 2001), however.  
All of these new detections have been confirmed with ATCA, and have M$_{HI}$= 
10$^{7-9}$\msun.  All of the detections have associated stellar components; 
they are all dwarf galaxies.  

\begin{figure}
\plotone{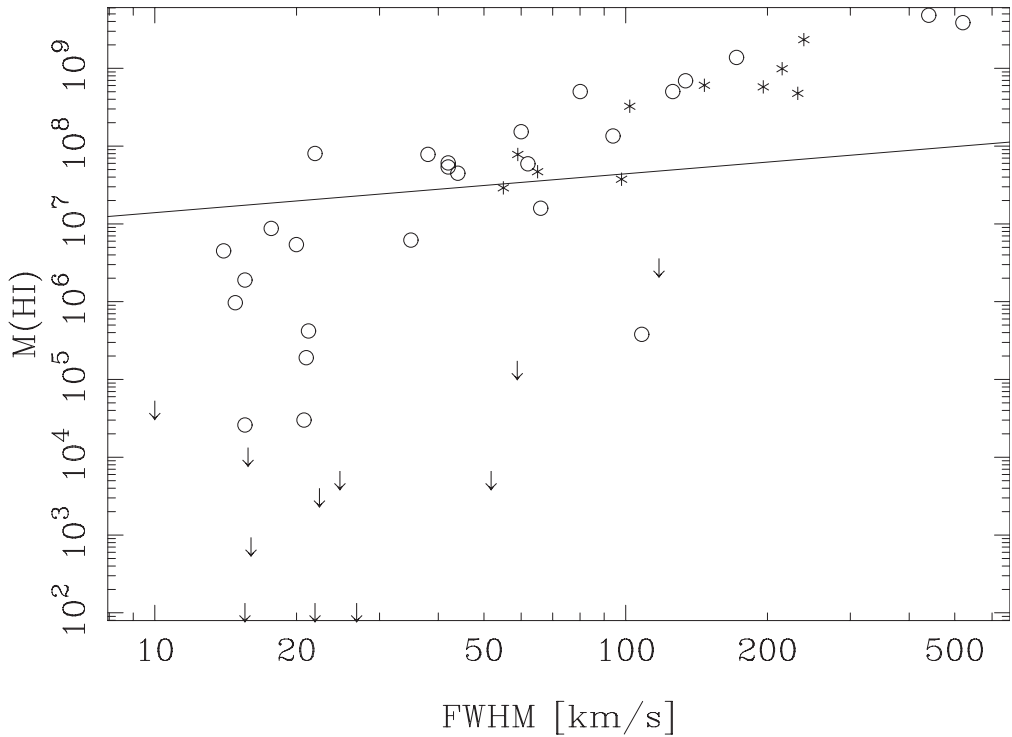}
\caption{\mhi\ vs. V$_{rot}$ for Local Group galaxies (from Mateo 1998) with 
measured \mhi\ (circles) and upper limits (arrows).  The stars indicate the
sources in LGG 93.  The line indicates the 10$\sigma$ detection limit}
\end{figure}

Our survey found that the ``missing satellite'' problem is actually worse in 
the groups we surveyed than in the Local Group.  Figure 1 shows that if we were
observing the Local Group, we would detect a total of 14 galaxies, 9 of which 
have FWHM $\le$100 \kms\ and could be considered dwarf galaxies.  Data for the 
Local Group galaxies comes from Mateo (1998).  The dwarf-to-giant ratio is 1.45
 for the three groups we studied, compared with 1.8 for the 
Local Group.  This discrepancy will be verified when we finish analyzing the 
remaining three groups in our sample, but, at the present, it appears that the
Local Group is not unique in its deficit of luminous satellite galaxies 
compared to cold dark matter models of galaxy formation.  

\section{The Nature of High-Velocity Clouds}

HVCs are clouds of \HI\ seen all around the Milky Way, but which
are not in regular galactic rotation (see Wakker, these proceedings for a 
review).  As such, their distances and masses are unknown, making their 
origins uncertain.  Some HVCs are certainly associated with the Magellanic 
Stream which is the result of the tidal interaction between the LMC, SMC, and 
Milky Way, and it is possible that other complexes have similar origins
(e.g. Lockman 2003).  Other HVCs may be ejected material from a Galactic 
fountain, or may be primordial gas falling into the Local Group and onto the
Milky Way as part of ongoing galaxy formation.  

While the idea of HVCs being associated with galaxy formation is an old one,
interest in the idea was revived by Blitz \etal\ (1999) and Braun \& Burton
(1999) who suggested that HVCs and compact HVCs (CHVCs) may contain dark 
matter, have masses of $\sim$10$^7$\msun, and reside at distances of $\sim$
1 Mpc from the Milky Way.  These HVCs may even be associated with the large
population of dark matter halos seen in the simulations discussed above.  
This idea has been expanded on by de Heij, Braun, \& Burton (2002b), who 
proposed that
CHVCs were concentrated around the Milky Way and M31 with a Gaussian distance
distribution of width 150 - 200 kpc, and \mhi\ $\le$10$^7$\msun.  If CHVCs 
are associated with galaxy formation, then some of their analogs should be 
visible in the groups we have observed.

Starting with this premise, we have constructed a simple model for HVCs to 
predict how many should be seen in the groups we have studied.  We begin by
taking the integrated fluxes and velocity widths of all Milky Way CHVCs 
cataloged by Putman \etal\ (2002) and de Heij, Braun, \& Burton (2002a).  We 
then randomly
determine a distance assuming a Gaussian distance distribution of a given 
half-width, half-maximum.  Given this distance, we get an \HI\ mass for each
cloud.  Now we ask if this CHVC were in one of our groups, would we detect
it at a 10$\sigma$ level.  We do this for every CHVC to determine the number
of expected detections per group for a given D$_{HWHM}$, and then we repeat the
process 100,000 times to get a distribution of expected detections.  The 
results of these trials for two groups, LGG 93 and LGG 478, and two D$_{HWHM}$ 
values are shown in Figure 2.  

\begin{figure}
\plottwo{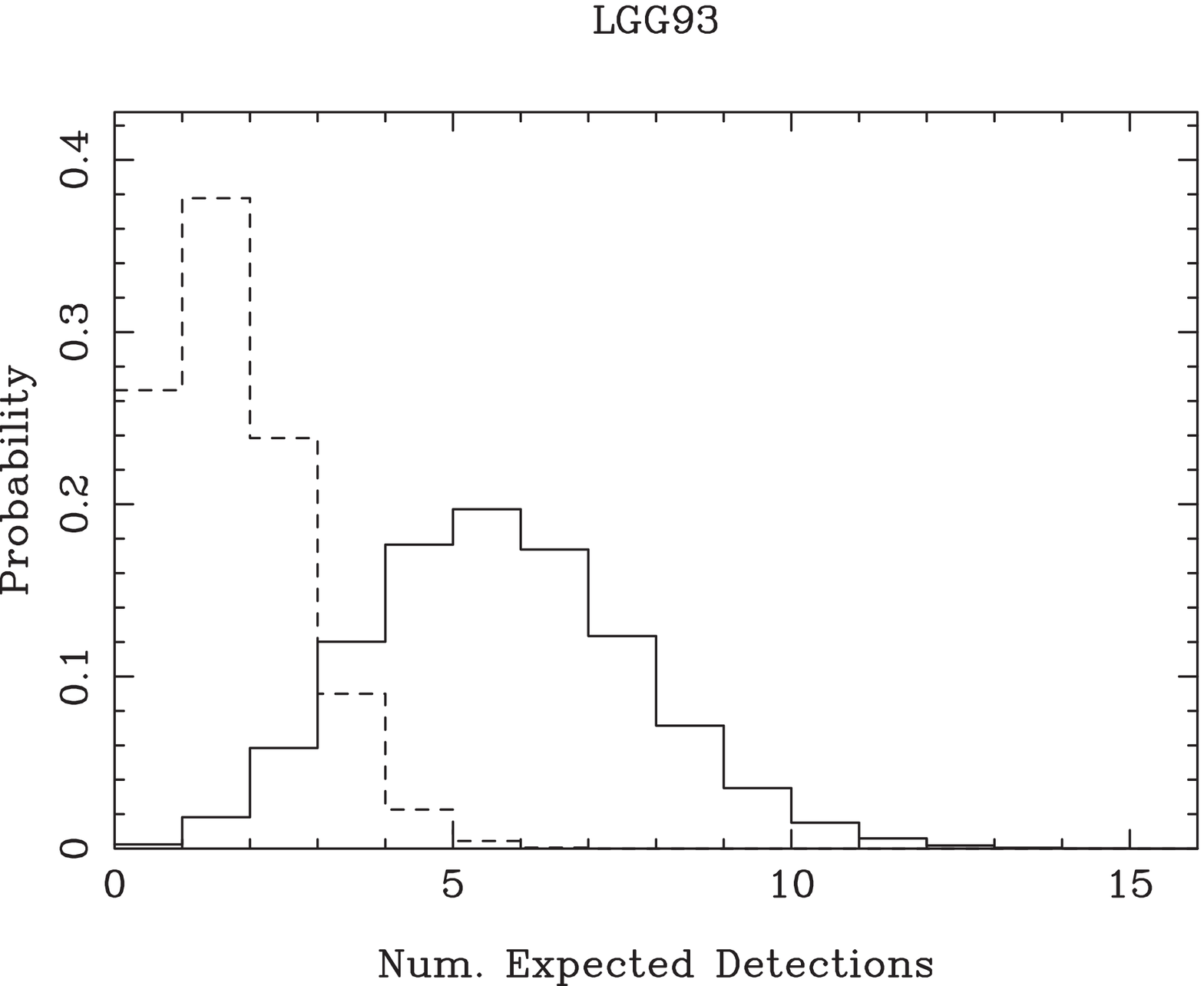}{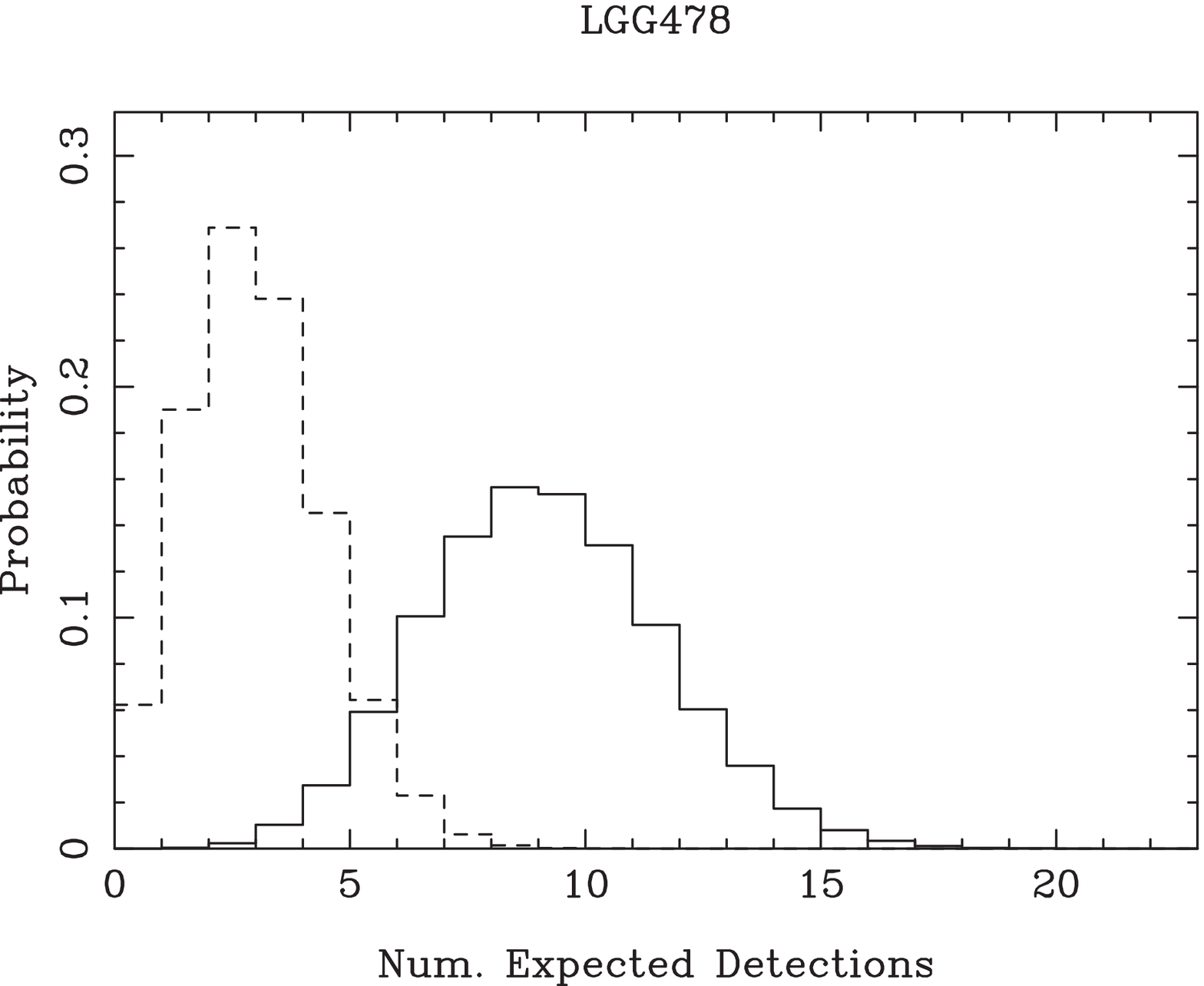}
\caption{The probability of a given number of detections for LGG 93 (left)
and LGG 478 (right).  The solid line represents the probability distribution
for D$_{HWHM}$ = 400 kpc and the dashed line represents D$_{HWHM}$ = 250 kpc.}
\end{figure}

As we already stated in Section 3, we have found no HVC analogs (\HI\ clouds
without stars) in any of our groups.  Therefore, in looking at Figure 2, we
want to know for a given D$_{HWHM}$ what the probability of zero detections 
is.  Since this is a function of the mass detection limit of the group, we
get different limits on D$_{HWHM}$ for each group.  For LGG 93, we can see
in Figure 2 that D$_{HWHM}$ is less than 400 kpc at a 99.76\% confidence
level, but D$_{HWHM} <$ 250 kpc is only 73\% likely.  For LGG 478, the figure
demonstrates that D$_{HWHM} <$ 400 kpc is 99.997\% likely, and D$_{HWHM} <$
250 kpc is 93.8\% certain.  And for LGG 180, not shown in the figure, the 
limits are 99.98\% and 89\% certain, respectively.  

There a couple of notes of caution regarding this analysis.  First, it is 
important to note that the vast majority of CHVC analogs are not detected; we 
can only detect the most massive analogs.  This means that we are using a 
small number of objects to infer the properties of a larger population.  If 
the population of CHVCs in these groups is different than in the Local Group,
we may not be able to detect any analogs.  We also assume that the same 
{\it number} of CHVCs is present in each group.  
If the number of CHVCs is a function of group mass, for example, we would 
expect less CHVCs in LGG 478 than in LGG 93 or LGG 180, so the constraints on
D$_{HWHM}$ may not be as strong as suggested above.  Nevertheless by analogy
to other groups, it appears that CHVCs should be clustered with D$_{HWHM} \le$
300-400 kpc of the Milky Way.  At these distances, the total \HI\ mass in 
CHVCs is only $\sim$10$^8$\msun, making them an important source of fuel for
star formation, but not dynamically important to the Local Group.  

\section{Conclusions}

We have surveyed $\sim$1 Mpc$^2$ around the centers of 3 loose groups of 
galaxies using the Parkes Multibeam and the ATCA.  The goal of these 
observations is to get a census of the \HI-rich galaxies in these groups and 
to search for analogs to HVCs.  This will permit us to constrain models for 
the origin of HVCs and
to begin to test models of galaxy formation.  These groups were chosen
to be analogs to the Local Group, so that they only contain a few large spiral 
galaxies which are separated by a few hundred kiloparsecs.  Our observations 
have very high velocity resolution, $<$3.3 km/s, in order to facilitate the
detection of low-mass objects, so that our 1$\sigma$ detection limit is
$\sim$7$\times$10$^5$\msun\ per 3.3 km/s.  Using fake sources to determine our 
completeness and ATCA observations to confirm our detections, we are roughly
100\% complete and reliable at 10$\times$ the theoretical noise limit.  

We found 7 new \HI-rich objects in the groups, plus 7 background galaxies, 
only 4 of which were not seen in HIPASS, and all of which have stellar 
counterparts.  The ratio of low-mass to 
high-mass galaxies in these groups is less than expected from simulations and 
also less than the ratio in the Local Group, therefore the ``missing 
satellite'' problem is not unique to our local neighborhood.  

We found no analogs to HVCs in any of the three groups.  Assuming the HVCs in 
these groups are the same as in the Local Group, this implies that compact HVCs
must be clustered within $\sim$300-400 kpc of the Milky Way, otherwise we would
have seen their analogs in our survey.  This is strong evidence against the 
Blitz \etal\ (1999) and Braun \& Burton (1999) models for the origin of HVCs, 
but is still consistent with the de Heij \etal\ (2002b) model.  At these 
distances, however, the total \HI\ mass in CHVCs is only $\sim$10$^8$\msun,
making them unimportant to the dynamics of the Local Group, but still a 
useful reservoir of fuel for future star formation.  

We have Parkes data in hand for an additional 3 groups, and will be following 
up our detections in these groups with the ATCA and the VLA in the near future.
The ensemble of the data on these 6 groups will place stronger constraints on 
the nature of HVCs and provide a better test of different models of galaxy 
formation.

\acknowledgements

The authors wish to thank the excellent staff at Parkes and the ATCA for their 
assistance with observing.  We wish to thank Martin Zwaan for his assistance 
with inserting fake sources into our data cubes.  D.J.P. acknowledges generous 
support from NSF MPS Distinguished Research Fellowship grant AST0104439.

\end{document}